\documentclass[reprint,amsmath,amssymb, aps,]{revtex4-1}
\usepackage{graphicx}
\usepackage{dcolumn}
\usepackage{bm}
\usepackage{hyperref}

\begin{document}
\title{Charge Fluctuations of an Unchareged Black Hole}
\author{Marcelo Schiffer}
\thanks{email:schiffer@ariel.ac.il}%
\affiliation{Department of Physics, Ariel University, Ariel 44837, Israel.}

\date{\today}

\begin{abstract}

In this paper we calculate  charge fluctuations of a Schwarzschild black-hole  of mass $M$ confined within a perfectly reflecting cavity of radius R in  thermal equilibrium with various species of radiation and  fermions . Charge conservation is constrained by a Lagrange multiplier (the chemical potential). Black hole charge fluctuations are expected owing  to continuous absorption and emission of particles by the black hole. For black holes much more massive than  $10^{16} g$ , these fluctuations are exponentially suppressed. For black holes lighter than this, the Schwarzschild black hole is unstable under charge fluctuations for almost every possible size of the confining vessel. The stability regime  and the fluctuations are calculated through the second derivative of the entropy with respect to the charge. The expression obtained contains many puzzling terms besides the expected thermodynamical fluctuations: terms corresponding to instabilities that do not depend on the specific value of charge of the charge carriers and  one of them depends on Newton's constant instead. One of the contributions to the charge fluctuations $\hbar/4\pi$ does not depend neither on number of species, nor on the the specific charge or even the size of the confining vessel.  As a matter of fact, this term emerges from the second derivative of the black hole entropy alone, which means that it corresponds to a  genuine quantum mechanical property of the black hole itself. Such a contribution would cause the event horizon to recede from $2M$ to $2M-T_{BH}$ or equivalently, by $(4\pi)^{-1}$ of the black hole' s Compton wave length. Similarly,  a Cauchy horizon emerges at the same distance the event horizon receded. 
\begin{description}
\item[PACS numbers : 04.70Dy,  05.40.-a,  05.70-a,52.25 Kn]
\end{description}
\end{abstract}

\maketitle

\section{Introduction}
An infalling  electron crosses the event horizon of an otherwise neutral massive black-hole. At the point it crosses the event horizon, it causes $r_+$ to recede by $\delta r_+=M-\sqrt{M^2-Q^2}$. The receded horizon  intersects the  black-hole  time-like singularity, leaving a sector of the singularity uncovered by an horizon,  clashing  with the cosmic censorship hypothesis which forbids  naked singularities. On the other hand, we know that the singularity of a charged black hole is space-like: at the point the particle reaches the singularity (point A) the singularity flips from time-like to space like, it is then entirely dressed by the the event horizon, the cosmic censorship hypothesis is spared. Furthermore, a Cauchy horizon $r_-$ is formed, closing the space-time diagram. 

\begin{figure}[htbp]
   \centering
   \includegraphics{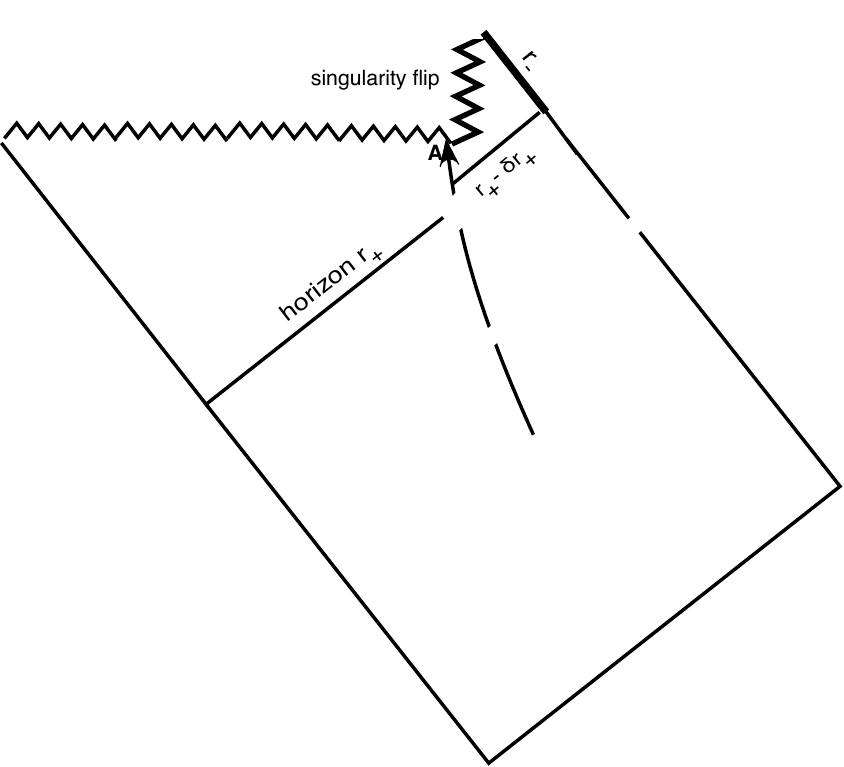} 
   \caption{topologic change}
   \label{fig:example}
\end{figure}

 Such a topological change of a very massive black hole caused by say, a single infalling electron seems disturbing. On the other hand, if charge fluctuations of an otherwise uncharged black hole are always present,  no black hole is ever   truly uncharged; these charge fluctuations prevent a topological swap.  In order to address the issue,  we resort to a controlled situation of an uncharged black hole in thermal equilibrium with a number $g_b$ species of  gauge  bosons and  $g_f $  species of  electrically charged fermion pairs. The black hole and the  plasma reside inside a perfectly reflecting spherical cavity of coordinate radius $R$; the overall  charge of the system, as well as the mean charge of the black-hole vanishes.  We tacitly assume  that black hole charge fluctuations   are produced by fluctuations of the pair production process at the event horizon (Hawking radiation) and the continuous absorption of particles by the black-hole. Gauge bosons are required for consistency, as pairs of charged particles are continuously  produced and annihilated within the plasma and one of the pair's member might be absorbed by the black hole, or otherwise, emitted  by the hole and annihilated by some charge unbalance in the plasma. 

\section{A Black Hole in Thermal Equilibrium with a Plasma}

Consider a black hole of  mass $M$ and charge  $Q$ in thermal equilibrium with a plasma consisting of $g_b$ gauge bosons and $g_f$  charged electron and positron pairs. The primordial quantity to assess the thermodynamical equilibrium is the total entropy:
\begin{equation}
S_{tot}=S_{bh} + S_\pm +S_\gamma  
\label{entropy}
\end{equation}
where $S_{bh}, S_\gamma$ and $S_\pm$ are  respectively the black hole, radiation and charged particles' entropies. Furthermore,  the energy and charge conservation constraints must be fulfied:
\begin{eqnarray}
E_{tot}&=&M + E_\pm +E_\gamma \\
\label{energy}
0&=& Q+e (N_+-N_-)
\label{charge}
\end{eqnarray}
where $E_\gamma$ and $E_\pm$ are the boson and fermion contributions to the the total energy, $e$ the charge of the fermions, taken identical for all species and $N_+$,$N_-$ the total number of positive and negative charged particles inside the cavity.

With the the appropriate number of particles per mode $n(\epsilon)$, we can calculate the total energy and the total number of particles:
\begin{eqnarray}
E&=&g \int d\Sigma^0 \int \frac{d^3p}{h^3} \int  \ \epsilon n(\epsilon) \nonumber\\
N&=&g \int d\Sigma^0 \int \frac{d^3p}{h^3} \int g(p)  n(\epsilon) \\
\end{eqnarray}
where $g$  is the degeneracy of each species and $ d\Sigma^0$ ,the spatial volume element of the time foliation.  Last, the black-hole entropy is given by Bekenstein's formula ~\cite{yakov} :
\begin{equation}
S_{BH} =\frac{1}{4\hbar} (4\pi r_+)^2
\end{equation}
where $r_+= M+\sqrt{M^2-Q^2}$.

Accordingly, the charge conservation constraint reads
  \begin{equation}
0=Q+ \frac{4 e g_f}{\pi \hbar^3} \int _{r_+}^R  r^2\sqrt{g_{rr}} dr \int  \left (n_+- n_-\right) p^2 dp
  \label{X}
 \end{equation} 
while  the energy condition reads
\begin{eqnarray}
 && E=M+ g_b \frac{4\pi^3}{15 \hbar^3}  \beta^{-4} \int _{r_+}^R  r^2\sqrt{g_{rr}} dr  + \nonumber\\
&& \frac{4g_f}{\pi \hbar^3}  \int _{r_+}^R r^2\sqrt{g_{rr}} dr \int\left[ \epsilon \left (n_++n_-\right)+
\frac{e Q}{r} (n_+-n_-) \right]p^2 dp \nonumber\\
 \label{E}
\end{eqnarray}
In the above expressions $\epsilon =\sqrt{p^2+m^2}$ with  $m$  the mass of the fermions, taken identical for all species , $R$ is the coordinate radius of a spherical confining vessel and $\pm eQ/r$ is the electrostatic energy for each positive and negative fermion. The  number of particles per mode is: 
\begin{equation}
 n_\pm=\frac{1}{e^{\beta(\epsilon \pm \nu)}+1}
 \end{equation} 
with
  \begin{equation}
  \nu =eQ \mu + \frac{eQ}{r}
  \label{nu}
  \end{equation}
Here $\mu$ stands for the chemical potential in units of $eQ$. Equilibrium considerations require the first the entropy derivatives with respect to the control parameters to vanish. Second derivatives provide information about stability and the amount of fluctuations.  Therefore all quantities should  be calculated to order not larger than $\mathcal{O}(Q^2)$, if we are concerned with fluctuations around an uncharged configuration. Thus, up to this order:
 
\begin{eqnarray}
n_\pm &=&\frac{1}{e^{\beta \epsilon}+1} \mp \frac{\beta \nu}{4}\frac{1}{ \cosh^2(\beta \epsilon/2)}+ \frac{(\beta \nu)^2}{8}\frac{\sinh(\beta \epsilon/2)}{\cosh^3(\beta \epsilon/2)}  +\dots \nonumber\\
& &
\end{eqnarray}
where we tacitely assumed that the chemical potential is  also linear in $Q$, $\mu \sim {\mathcal O}(Q^0)$ or equivalently $\nu \sim \mathcal{O}(Q)$.  Under these conditions, we can immediately solve the charge conservation constraint [eq.(\ref{X})]:
\begin{equation}
\mu= \frac{3\hbar^3 \beta^2}{4\pi e^2 g_f AI_2}-\frac{I_1}{I_2}
\label{mu}
\end{equation}
where we defined the spatial integrals 
\begin{equation}
I_n = \int_{2M}^R \frac{r^n}{\sqrt{1-\frac{2M}{r}}}dr 
\label{sti}
\end{equation}
and 
\begin{equation}
A(x)=\frac{12}{\pi^2} \int_0^\infty \frac{q^2 }{\cosh^2 \xi} dq 
\end{equation}
where $\xi \equiv \sqrt{q^2 +(x/2)^2}$  with $x=m\beta$ ; this definition of $A(x)$ and the other functions that follow was conveniently chosen  such that in the relativistic limit $x \rightarrow 0 $ all these functions approach one.  
Since eq.(\ref{mu}) does not depends on $Q$, our assumption that $\mu \sim {\mathcal O}(Q^0)$ is completely justified. The energy constraint looks a bit more cumbersome 
\begin{eqnarray}
E&=&M +\frac{\pi^3 \beta^{-4}}{15  \hbar^3} \left(4g_b+7g_f B  \right)  \int _{r_+}^R r^2\sqrt{g_{rr}} dr\nonumber\\
&+&\frac{2\pi g_f e^2Q^2 \beta^{-2}}{3\hbar^3} \left[ 3C( \mu^2  I_2 +2\mu I_1 +I_0) -2A(\mu I_1 +I_0) \right] \nonumber\\
\end{eqnarray}
where we defined two new integrals
\begin{eqnarray}
B(x)&=&\frac{120}{7 \pi^4} \int_0^\infty \frac{\chi q^2 }{e^\chi+1} dq\\
C(x)&=&\frac{8}{\pi^2} \int_0^\infty \xi q^2 \frac{ \sinh \xi }{\cosh^3 \xi} dq
\end{eqnarray}
with $\chi=\sqrt{q^2+x^2}$. Both integrals are to be evaluated for $x=m\beta$;  $B(x)\rightarrow 1,C(x)\rightarrow 1$ in the relativistic limit. 

The radiation entropy is obtained via the first law of thermodynamics;  the  fermion's entropy can be calculated in this same manner, provided we implement the  charge conservation constraint . It is less cumbersome, however to calculate this entropy from its definition: 
\begin{widetext}
\begin{equation}
S_f=g\int  \sqrt{-g}\frac{d^3rd^3p}{h^3} \left[
\frac{\beta(\epsilon + eQ/r)+e Q \mu}{e^{\beta((\epsilon +eQ/r)+eQ\mu)}+1} +\log\left(1+e^{-\beta(\epsilon+e Q/r)+eQ\mu)}\right)
\right]
\end{equation}
As  the standard procedure, we integrate  the second term in this expression by parts  and write for the total entropy
\begin{equation}
S=\frac{\pi}{\hbar} (M+\sqrt{M^2-Q^2})^2+ \frac{16\pi^3 g_b\beta^{-3}}{45  \hbar^3}\int_ {r_+}^R r^2\sqrt{g_{rr}} dr
+ \frac{4 g_f \beta}{\pi \hbar^3} \int_{r_+}^R\sqrt{g_{rr}}r^2 dr  \int  \left[(\epsilon +\frac{p^2}{3\epsilon})(n_++n_-)+\nu(n_+-n_-)\right]p^2dp
\end{equation}
where we used the fact that $\epsilon^2=p^2+m^2$. Up to the second order in $Q$:
\begin{eqnarray}
S&=& \frac{2\pi}{\hbar} (2M^2-Q^2) + \frac{\pi^3 \beta^{-3}}{45  \hbar^3} \left(16g_b+7g_f( 3B+E)\right)  \int _{r_+}^R r^2\sqrt{g_{rr}} dr\nonumber\\
&+& \frac{2\pi g_f e^2Q^2  \beta^{-1}}{3\hbar^3} (3C+D-2A)\left(I_2 \mu^2 +2\mu I_1+I_0\right)
\end{eqnarray}
\end{widetext}
where we defined two additional integrals
\begin{eqnarray}
D(x)&=&\frac{8}{\pi^2} \int_0^\infty \frac{q^4 \sinh \xi}{ \xi\cosh^3 \xi} dq\\
E(x)&=&\frac{120}{7\pi^4} \int_0^\infty \frac{1}{e^\chi+1}\frac{q^4 dq}{\chi }
\end{eqnarray}
with the same property that $D, E\rightarrow 1$ in the relativistic limit. 
To the leading term in $Q$, the volume integral reads
\begin{equation}
\int_{r_+}^R \sqrt{g_{rr}}r^2 dr= I_2 +\left(\frac{Q}{2M}\right)^2 \left[ \frac{R^3}{\sqrt{1-2M/R}}- 3 I_2+M I_1\right]
\end{equation}
Collecting all the terms, we can write the total energy and entropy in powers of $Q^2$ too:
\begin{eqnarray}
E&=&E_0 + Q^2 E_2\\
S&=&S_0+Q^2 S_2 
\end{eqnarray}
where
\begin{equation}
E_0=M +\frac{\pi^3 \beta^{-4}}{15  \hbar^3} \left(4g_b+7g_f B \right)I_2 
\label{E0}
\end{equation}
\begin{equation}
S_0= \frac{4\pi M^2}{\hbar}  + \frac{\pi^3 \beta^{-3}}{45  \hbar^3} \left[16g_b+7g_f(E+3B)\right]  I_2 
\label{S0}
\end{equation}
are the zero order terms. The second order terms are
\begin{widetext}
\begin{eqnarray}
E_2=\frac{\pi^3 \beta^{-4}}{60M^2 \hbar^3} \left(4g_b+7g_f B\right) \left( \frac{R^3}{\sqrt{1-2M/R}}- 3 I_2+M I_1\right)
+\frac{2\pi g_f e^2 \beta^{-2}}{3\hbar^3} \left[3 C\left(\mu^2 I_2+2\mu I_1 +I_0\right) -
2A(\mu I_1+I_0)\right] \nonumber\\
\end{eqnarray}
\begin{eqnarray}
S_2=-\frac{2\pi}{\hbar} + \frac{\pi^3 \beta^{-3}}{180M^2  \hbar^3} \left(16g_b+7g_f(E+3B)\right)  \left( \frac{R^3}{\sqrt{1-2M/R}}- 3 I_2+M I_1\right)
 + \frac{ 2\pi g_f e^2  \beta^{-1}} {3\hbar^3} \left(3C+D-2A\right)\left(I_2 \mu^2 +2\mu I_1+I_0\right) \nonumber\\
 \label{S2}
\end{eqnarray}
\end{widetext}

For a Schwarzschild  bh ($Q\rightarrow0$) in equilibrium with  the thermal bath of radiation and  plasma the  first  and second derivatives of the total entropy with respect to $M$ produce the usual thermal equilibrium conditions for a Schwzschild black hole with its radiation which was studied in detail  by  Pavon and many  others a long time ago ~\cite{pavon1}-~\cite{pavon2}. The first derivative of the total entropy with respect to the charge vanishes identically as it should be: an uncharged  black hole is an equilibrium configuration with its radiation.  Is is a stable configuration ? What is the size of the charge fluctuations ?  In order to answer these questions we proceed with the calculation of the second derivative of the entropy. The charge conservation constraint was already solved. Thus, the energy conservation constraint requires that:
\begin{eqnarray}
\left(\frac{\partial^2 E}{\partial Q^2}\right)_{Q=0}&=&2 E_2 +\frac{\partial E_0}{\partial \beta}\left(\frac{\partial^2 \beta}{\partial Q^2}\right)_{Q=0} \\  \nonumber &+&\left(\frac{\partial \beta }{\partial Q}\right)^2 \left(\frac{\partial^2 E_0}{\partial \beta^2}\right)_{Q=0}  =0
\label{E2}
\end{eqnarray}

Having in mind that  $\beta=\beta(Q^2)$, then  $(\partial \beta/\partial Q)_{Q=0}=0$  and it follows that
\begin{equation}
\left(\frac{\partial^2 \beta}{\partial Q^2}\right)_{Q=0} =- 2 E_2 \left(\frac{\partial E_0}{\partial \beta}\right)^{-1}
\end{equation}
On similar grounds
\begin{equation}
\left(\frac{\partial^2 S}{\partial Q^2}\right)_{Q=0}=2 S_2 +\frac{\partial S_0}{\partial \beta}\left(\frac{\partial^2 \beta}{\partial Q^2}\right)_{Q=0}  \label{S2}
\end{equation}
Putting these pieces together
\begin{equation}
\left(\frac{\partial^2 S}{\partial Q^2}\right)_{Q=0}=2\left(S_2 - X \beta E_2\right)
\label{almosthere}
\end{equation}
where we defined
\begin{equation}
X= \beta^{-1}\frac{(\partial S_0/\partial \beta)}{(\partial E_0/\partial \beta)}
\end{equation}
Having in mind the definitions of $E_0$ [eq.(\ref{E0})]and $S_0$ [eq.(\ref{S0})] we have explicitly 
\begin{equation}
X(x)=\frac{16 g_B +7(3B(x)+E(x) )g_f -7g_f x(B'(x)+E'(x)/3)}{16g_b +28 B(x)g_f -7g_f x B'(x)}
\end{equation}
In the ultra relativistic limit,  $x\rightarrow 0$ all the above  functions in the definition of $X(x)$ approach one while their derivatives vanish and, accordingly $X\rightarrow 1$. In the opposite low temperature limit, all these integrals vanish exponentially,  and again $X\rightarrow 1$ .  We plotted the graph of the function $X(x)$  for one species of gauge bosons and one of charged particle/antiparticle  fermion. 
\begin{figure}[htbp]
   \centering
   \includegraphics[width=0.4\linewidth, height=0.15\textheight]{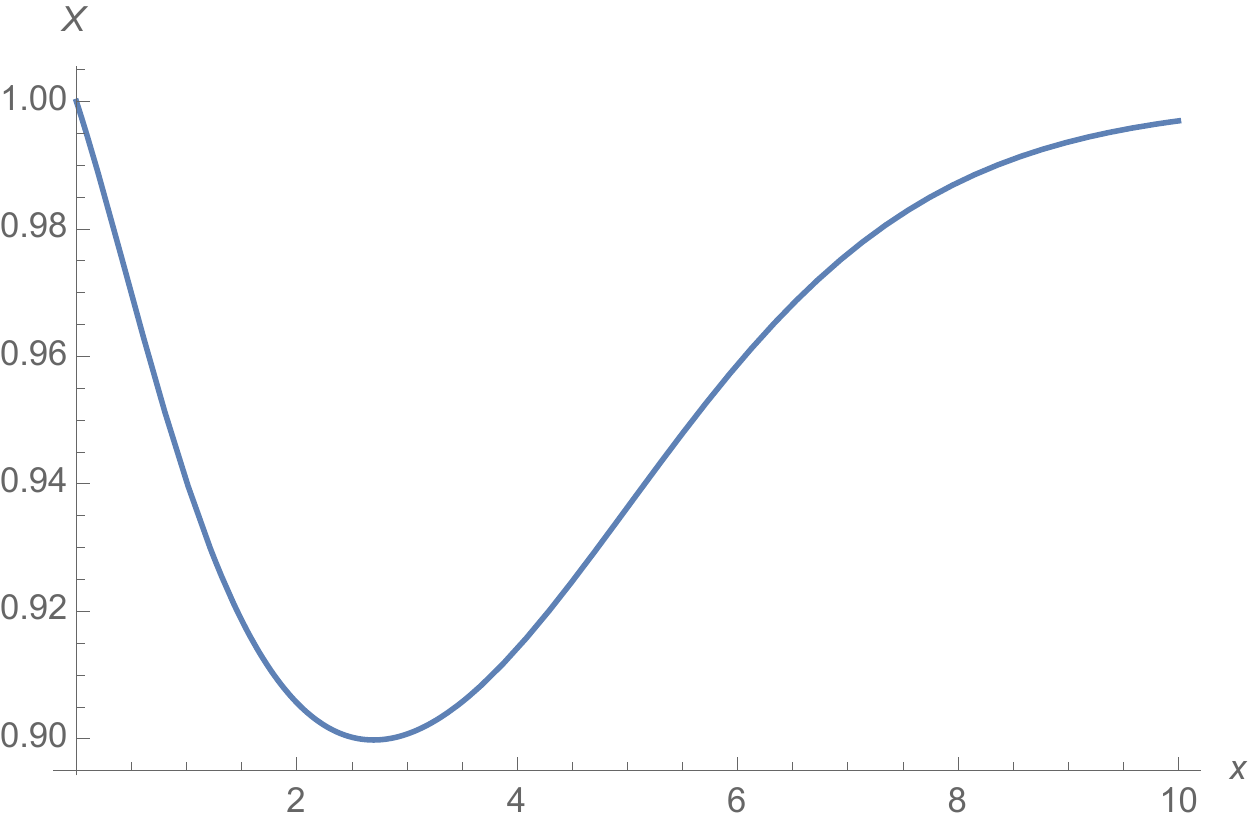} 
   \caption{The function $X(x)$ for $g_f=g_b=1$}
   \label{fig:example}
\end{figure}
At any rate, $X(x)\leq 0.9$. 

\begin{figure}[htbp]
   \centering
   \includegraphics[width=0.4\linewidth, height=0.15\textheight]{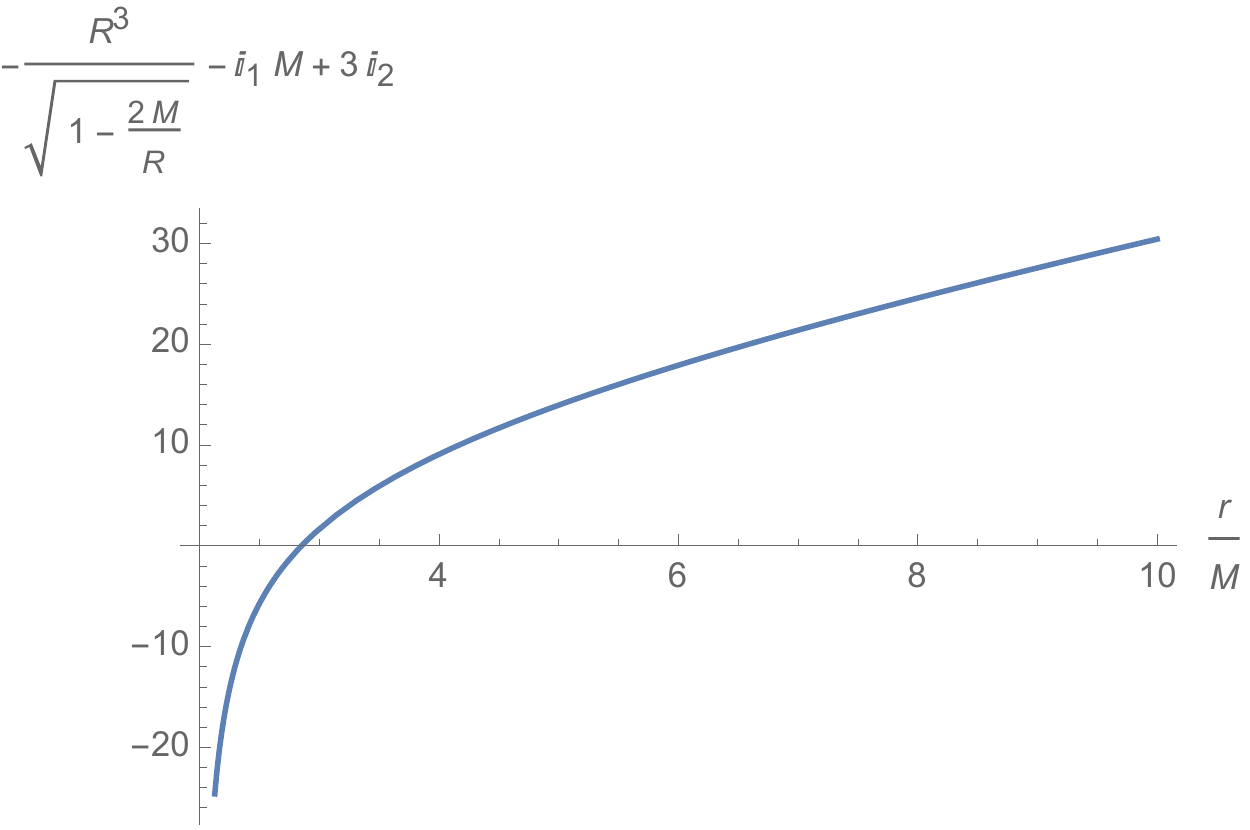}
   \caption{The function is positive beyond the instability radius $r\geq 3M$}
   \label{I3}
\end{figure}

Recalling the expressions for $E_2$ [eq.(\ref{E2})] , $S_2$ [eq.(\ref{S2})] and finally   for $\mu$  [eq.(\ref{mu})] , it follows that:
\begin{widetext}
\begin{equation}
\begin{split}
\left(\frac{\partial^2 S}{\partial Q^2}\right)_{Q=0}=-\frac{4\pi}{\hbar}-\frac{\pi^3}{90M^2 \hbar^3 \beta^3 G}\left[4g_b(4-3X) + 7g_f 
\left(E +3B(1-X)\right)\right] \left(3 I_2 -\frac{R^3}{\sqrt{1-\frac{2M}{R}}} -MR_1\right) \\
-\frac{3 \hbar^3 \beta^3}{4\pi  e^2 g_f I_2 } \left[\frac{2 A -D -3C(1-X)}{A^2}\right] +\frac{4\pi g_f e^2}{3\beta \hbar^3}\left[(3C-2A)(1-X)+D\right] \left(I_0-\frac{I_1^2}{I_2}\right)+2\beta X\frac{I_1}{I_2} 
\end{split}
\label{SQQ}
 \end{equation}
\end{widetext}
where  Newton's constant was returned in place.  We checked numerically that all the combinations of the functions $A(x),...,E(x)$ in the above expressions are positive definite. The function  $I_0-I_1/I_2^2$ is positive definite while  inspection of fig.(\ref{I3})  
reveals that $3I_3- R^3/\sqrt{1-2MG/R}-M I_1 $ is positive for $R> 2.85M $, which curiously is very close to the last stable photon orbit  \cite{MTW}.  Clearly we need to require that the confining vessel is larger than those unstable orbits, in which case this function is also positive. The three first terms contribute to the thermodynamical stability of the $Q=0$ configuration , while the last two contribute to the instability. Let us know consider a very large confining cavity $R>>2M$, in which case
\begin{equation}
3I_3-\frac{R^3}{\sqrt{1-2MG/R}}-M I_1 \approx  2 M^2 G^2 R \quad , \nonumber
\end{equation}
\vspace{2pt}
\begin{equation}
I_0-\frac{I_1^2}{I_2} \approx \frac{R}{4}  \quad , \quad  \frac{I_1}{I_2} \approx \frac{3}{2R}
\end{equation}
and further taking  the relativistic limit $m\beta <<1$,  we have for the charge fluctuations
\begin{widetext}
\begin{equation}
\begin{split}
\frac{1}{\Delta Q^2}
=\frac{4\pi}{\hbar}+\frac{(4g_b+7g_f)\pi^3}{45} \frac{R G} {\hbar^3 \beta^3} +
\frac{9 }{4\pi g_f }\frac{\hbar ^3 \beta^3 }{ e^2 R^3} -\frac{\pi g_f }{3}\frac{e^2 R}{\beta \hbar^3}-3 \frac{\beta}{R} 
\end{split}
\label{unstable}
 \end{equation}
\end{widetext}
We can identify 
\begin{equation}
\Delta Q^2_{td}=\frac{4\pi g_f }{9 }\frac{ e^2 R^3}{\hbar ^3 \beta^3 }
\end{equation}
as the thermodynamical charge fluctuations inside of the confining cavity, which must be identical to those of the black hole itself.  Interpretation of all the other terms remains elusive. For instance, the second term in the sum does not depend on the specific value of the charge of the charge carriers, neither does the last term which contributes for instability.  Recalling that $\beta \sim MG/\hbar$, the ratio of the two linear terms in $R$ is  of the order 
\begin{equation}
\frac{G}{e^2 \beta^2} \sim \frac{\left(M_p/M\right)^2}{\left(e^2/\hbar\right)} << 1 
\end{equation}
while the ratio of the two last terms is
\begin{equation}
\frac{e^2 R^2}{\beta^2 \hbar^3} \sim \frac{e^2}{\hbar}\left(\frac{R}{MG}\right)^2
\end{equation}
This means that for radii $R/MG \stackrel{<}{\sim}12 $ the second term controls the instability and is much more significant than the second term.  On the other hand, the ratio of the third term (the one of thermodynamical origin) with 
the forth term   (that controls the instability) is of the order
\begin{equation}
\frac{\hbar^6 \beta^4}{e^4 R^4} \sim \frac{\hbar^2}{e^4}\left( \frac{MG}{R}\right)^4
\end{equation}
which tells that the for stability the cavity cannot be larger than a few times the Schwarzschild radius. Under these conditions, the black hole is unstable under charge fluctuations.  Clearly, this discussion overlooks the the role of charge fluctuations at the walls of the cavity, which could eventually  stabilize the system.  As long as $m\beta \sim 1$ similar conclusion applies,  as only the numerical coefficients in eq.(\ref{unstable}) change .  Accordingly,   instability results for any   black hole lighter than 
\begin{equation}
M \sim \frac{M_p}{m_e} M_p \sim 10^{16}g
\end{equation}
The first term  in eq.(\ref{unstable}) is the most intriguing one:
\begin{equation}
\Delta Q^2_{BH}=\frac{\hbar}{4\pi} \quad ,
\end{equation}
as it neither  depends on the size of the confining cavity, nor on the number of species  of particles, nor not even on the specific value of the charge of the charged particles. It is a genuine quantum mechanical property of the black hole. As a matter of fact, it arises from the second derivative  of $S_{BH}$ with respect to the charge.  This contribution is bound to remain for a radiating  black hole radiating isolated .  Note that this genuine quantum mechanical effect  implies on a recession of  the event horizon and the emergence of  Cauchy surface :
\begin{eqnarray}
\langle r_+\rangle &=&2M-\frac{\Delta Q^2}{2M} =2M-T_{BH}\\
\langle r_-\rangle &=& \frac{\Delta Q^2}{2M}=T_{BH}
\end{eqnarray}
Furthermore, this intrinsic black hole contribution to its own charge fluctuation answers the question we posed in the introduction, no black hole is ever truly uncharged, it produces charge fluctuations that maintains the topology  unchanged by the absorption of charged particles. The black hole entropy is such  that the hole is protected against  topological changes . 

What is the fate of a large black hole? If $m\beta >>1$   all the functions $A,B,....$ die away exponentially . Therefore in the expression  for the second derivative of the entropy [eq. (\ref{SQQ})], the term which is proportional to $\beta^3$ is negative and diverges exponentially. Therefore, the black hole becomes stable and the charge fluctuations become exponentially small. 

Almost forty years ago, in an unnoticed paper Bekenstein \cite{spread}  studied the mass fluctuations of Schwarzschild black hole and concluded that the standard squared deviation of the black hole mass becomes negative for black holes of masses larger than $M_{max} \sim 10^{11}g - 10^{15}g$ . He called this the width paradox.  Black hole fluctuations has been overlooked over the years and they might lead us to new insights into the nature of Quantum Theory and Gravity.

\begin{acknowledgments}
 I am thankful that my ways  become intertwined with those of  J. D. Bekenstein Z"L . I was among those who had the privilege of profiting from his brilliant mind and modesty, his sharpness, criticism and friendship. A man in search of the True and Righteousness.
 \end{acknowledgments}

 \begin{quotation}
\flushleft
\bibitem{hawkingellis}  {Hawking, S. W.; Ellis, G. F. R:  The Large Scale Structure of Space-Time. New York: Cambridge University Press  (1973).}
\bibitem{yakov}  {Bekenstein, Jacob D.. "Black holes and entropy". Physical Review D 7 (8): 2333, 2346. (1973).}
\bibitem{pavon1}{General Relativity and Gravitation ,D. Pavon, W. Israel , Vol. 16, Issue 6, pp 563-568 (1984).}
\bibitem{parentani}{R Parentani, J Katz, I Okamoto,  "Thermodynamics of a black hole in a cavity Classical and Quantum Gravity" - iopscience.iop.org (1995).}
\bibitem{pavon2}{D Pavon, J.M. Rubi,  " On some properties of the entropy of a system containing a black hole, General relativity and gravitation,  - Springer (1985).}
 \bibitem{hawking} {Hawking, S. W. (2005). "Information Loss in Black Holes ". arxiv.org/pdf/hepth/0507171.}
 \bibitem{MTW} C.W. Misner ; K.S. Thorne and J.A. Wheeler in {\it Gravitation}: Freeman 1973.
 \bibitem{spread} J.D. Bekenstein {\it Gravitation, the Quantum, and Statistical Physics}, in {\it To fulfill a Vision,} ed. Yuval Ne'eman , Addison Wesley: 1979.
\end{quotation}\noindent

\end{document}